\documentclass[twocolumn,pra,notitlepage]{revtex4-1}
\usepackage{amsmath}
\usepackage{graphicx}
\usepackage{wrapfig}
\usepackage{enumerate}
\usepackage{float}
\usepackage{standalone}

\begin{document}

\title{Polarization-independent optical spatial differentiation\\with a doubly-resonant one-dimensional guided-mode grating}

\author{Ali Akbar Darki$^1$, S{\o}ren Peder Madsen$^2$ and Aur\'{e}lien Dantan$^1$}\email{dantan@phys.au.dk}

\address{$^1$Department of Physics and Astronomy, Aarhus University, DK-8000 Aarhus C, Denmark\\
$^2$Department of Mechanical and Production Engineering, Aarhus University, DK-8000, Aarhus C, Denmark}

\begin{abstract}
We report on the design and experimental characterization of a suspended silicon nitride subwavelength grating possessing a polarization-independent guided-mode resonance at oblique incidence. At this resonant wavelength we observe that the transverse intensity profile of the transmitted beam is consistent with a first-order spatial differentiation of the incident beam profile in the direction of the grating periodicity, regardless of the incident light polarization. These observations are corroborated by full numerical simulations. The simple one-dimensional and symmetric design, combined with the thinness and excellent mechanical properties of these essentially loss-free dieletric films, is attractive for applications in optical processing, sensing and optomechanics.
\end{abstract}

\date{\today}

\maketitle


\section{Introduction}

Guided-mode resonant gratings are used within many applications in photonics and sensing~\cite{Wang1993,ChangHasnain2012,Quaranta2018,Cheben2018,Zhou2020}, e.g. in refractive index and biosensors, solar cells, photodetectors, spectrometers, lasers, etc. Recently, these gratings have also attracted a lot of interest for their applications within analog optical processing and computing~\cite{Zhou2020,Cheng2021}, where the possibility of realizing ultracompact, efficient and possibly reconfigurable optical components, such as lenses, waveplates, couplers, filters, etc. is exploited. 

Among the basic operations required for the processing of optical signals are the spatial differentiation and integration of optical beams~\cite{Cheng2021}. These have been investigated with metasurfaces~\cite{Silva2014,Pors2015,Zhu2021,Long2021} and resonant multilayered, plasmonic or photonic crystal structures~\cite{Golovastikov2014,Bykov2014,Ruan2015,Youssefi2016,Hwang2016,Zhu2017,Zangeneh2017,Zangeneh2018,Hwang2018,Guo2018}. Spatial differentiation with guided-mode resonant gratings in particular was proposed in~\cite{Golovastikov2014}, and recently experimentally demonstrated with TiO2 on quartz gratings~\cite{Bykov2018}, Si on quartz high-contrast gratings~\cite{Dong2018}, plasmonic gold subwavelength gratings~\cite{Yang2020}, suspended Si$_3$N$_4$ gratings~\cite{Parthenopoulos2021} and bilayer plasmonic gratings~\cite{Xu2021}. 

\begin{figure}[h]
\includegraphics[width=\columnwidth]{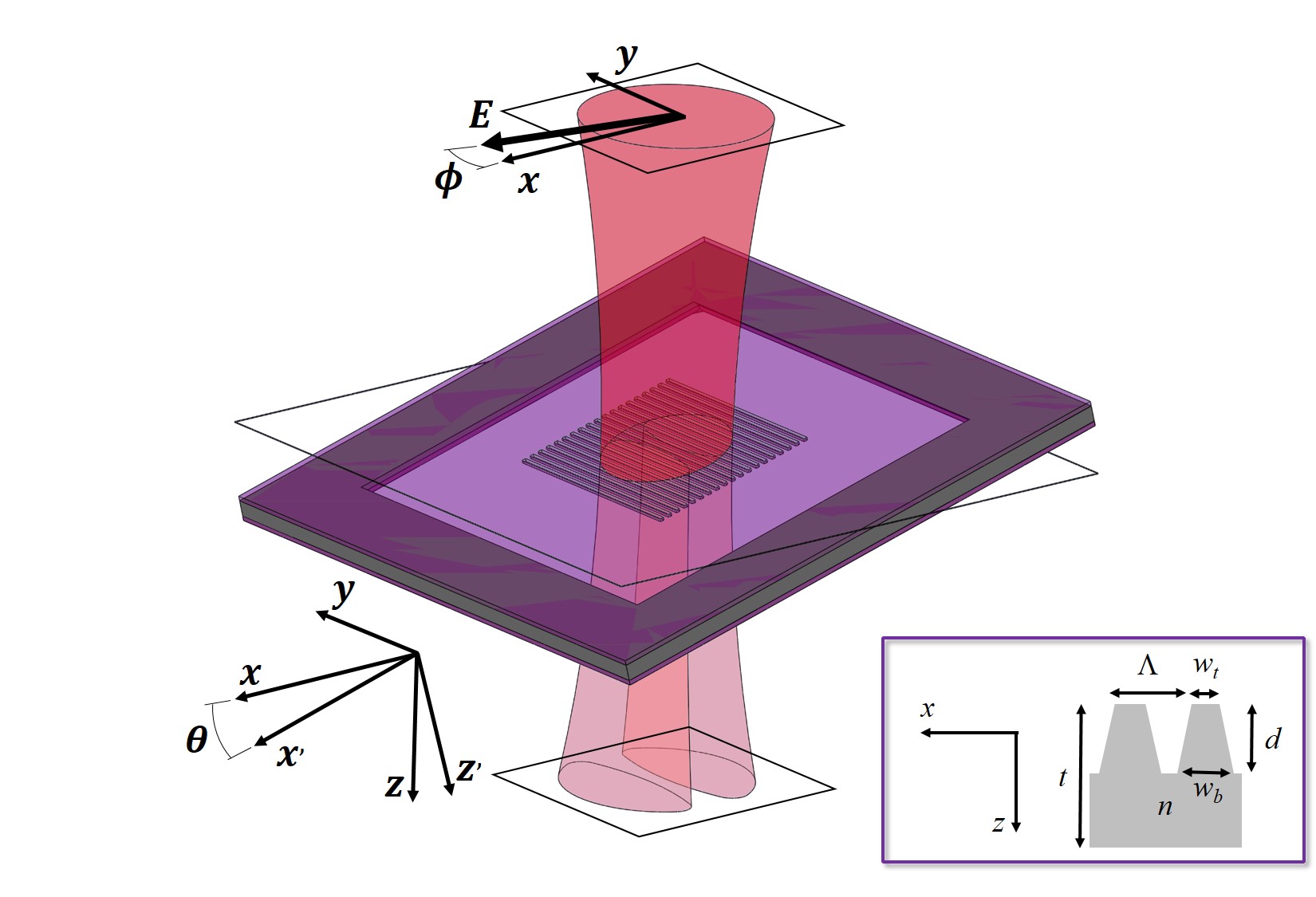}
\caption{Illustration of the situation considered: an optical beam, linearly polarized in the $(x,y)$-plane, impinges at oblique incidence on a suspended subwavelength grating, periodic in the $x$-direction and designed to possess a polarization-independent guided-mode resonance at this angle of incidence. At this resonant wavelength the transverse spatial profile of the transmitted beam (Gaussian, here) is differentiated at first-order in the direction of the grating periodicity, regardless of the incident light polarization (for clarity the reflected beam is not represented). The inset shows the geometry of the grating transverse structure.}
\label{fig1}
\end{figure}

In these realizations, first-order---as well as second-order in~\cite{Parthenopoulos2021}---differentiation of light with a specific polarization is observed around a specific wavelength, and, generally, the resonance conditions for the differentiation of TE- and TM-polarized light are widely different (see e.g. \cite{Parthenopoulos2021}). This naturally raises the question of whether polarization-independent spatial differentiators can be realized in such one-dimensional, symmetric resonant grating structures. Indeed, a canonical one-dimensional grating structure consists in a planar slab with a subwavelength periodic refractive index modulation---typically surrounded by a substrate and/or cover layers---in which the interaction between the light impinging at normal or oblique incidence on the grating and the light transversely propagating in the structure leads to resonant interference effects and the appearance of Fano resonances in the transmission or reflection spectra~\cite{Wang1993}. For a {\it homogeneous} slab the difference in the effective refractive indices for TE- and TM-polarized light prevents the TE- and TM-polarized guided modes supported by the slab from having the same propagation constant. However, this is no longer true in a periodically modulated structure, in which leaky modes with different polarizations can be made to simultaneously resonate, thereby achieving polarization-independent resonant response~\cite{Ding2004}. With a suitable design of the structure such a mechanism can be exploited for realizing e.g. polarization-independent guided-mode filters~\cite{Ding2004,Hu2010,Yangz2020,Sun2020}, whose fabrication may be simpler than two-dimensional grating~\cite{Mace2018}, cross-stacked grating~\cite{Kintaka2012,Kawanishi2019}, double-period grating~\cite{Sentenac2005,Fehrembach2007} or bilayered grating structures~\cite{Xu2021}. Let us also stress that, while various designs of such doubly-resonant one-dimensional gratings have been proposed, their experimental realizations are scarse and,  to the best of our knowledge, none has yet addressed their application to spatial differentiation.

Here, we design and demonstrate a polarization-independent, one-dimensional dielectric guided-mode grating allowing for polarization-independent, first-order optical spatial differentiation of optical beams and operating at oblique incidence and in transmission (see Fig.~\ref{fig1}). We show that, by patterning a 315 nm-thick suspended SiN film with a 200 $\mu$m-square, symmetric grating with 684 nm-period, 240 nm-deep fingers and 0.57-fill duty cycle, high-reflectivity guided-mode resonances can be simultaneously achieved at $\sim982$ nm for TM- and TE-polarized light impinging on the grating at a small angle of incidence ($3.6^\circ$). The measurement of the transmitted beam intensity profiles at resonance are consistent with those expected from first-order spatial differentiation of a Gaussian beam. The experimental observations are corroborated by predictions made on the basis of Rigorous Coupled Wave Analysis (RCWA) and Finite Element (FEM) simulations.

As abovementioned, such a simple one-dimensional grating design may ease the fabrication of polarization-independent guided-mode resonant grating-based devices. Furthermore, in addition to their excellent optical properties, these suspended nanostructured films show high mechanical quality with high-Q ($\sim$half a million) mechanical resonances in the MHz range, which make them well-suited for applications within sensing~\cite{Moura2018,Naesby2018,Dantan2020}, for optomechanics applications, e.g. for membrane-in-the-middle or membrane-at-the-end cavity optomechanics experiments~\cite{Kemiktarak2012,Bui2012,Kemiktarak2012a,Norte2016,Reinhardt2016,Chen2017,Cernotik2019}, for multi-membrane optomechanics investigations~\cite{Xuereb2012,Nair2017,Gartner2018,Piergentili2018,Wei2019,Yang2020b,Manjeshwar2020,Fitzgerald2021}. 

The paper is organized as follows: first, the design and simulations of the doubly-resonant guided-mode grating is presented in Sec.~\ref{sec:design}. In Sec.~\ref{sec:rcwa} the transmission spectra of the infinite structure under plane-wave illumination is analyzed on the basis of RCWA simulations. FEM simulations of the finite grating under Gaussian beam illumination are then carried out in Sec.~\ref{sec:fem}, allowing to predict and assess the spatial differentiation of the beam close to resonance. The experimental methods and results are then presented in Sec.~\ref{sec:exp}, before we conclude in Sec.~\ref{sec:conclusion}.


\section{Design and simulations}\label{sec:design}

\subsection{Doubly resonant guide-mode gratings}\label{sec:rcwa}

\begin{figure}[h]
\includegraphics[width=\columnwidth]{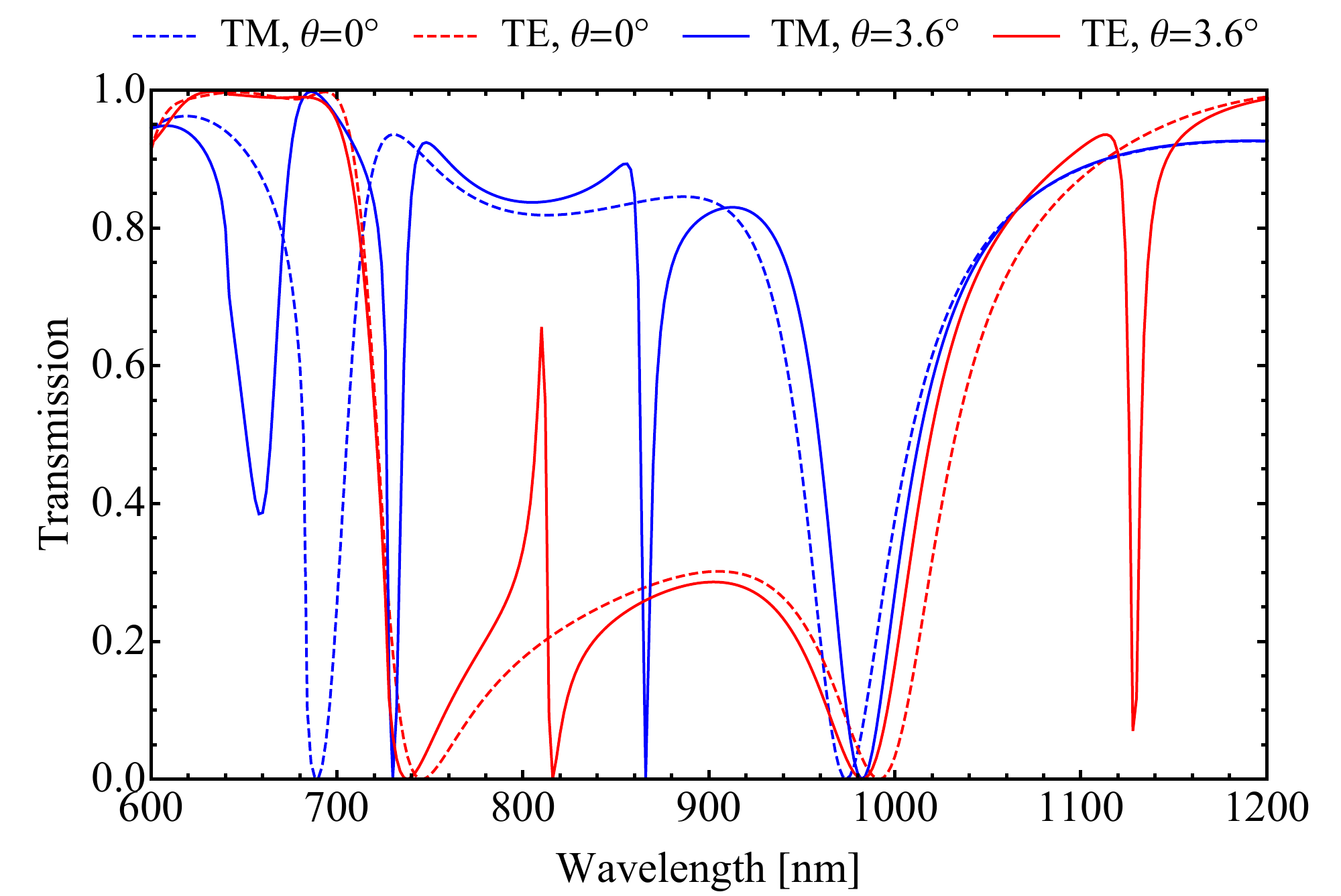}\\\vspace{0.4cm}
\includegraphics[width=0.97\columnwidth]{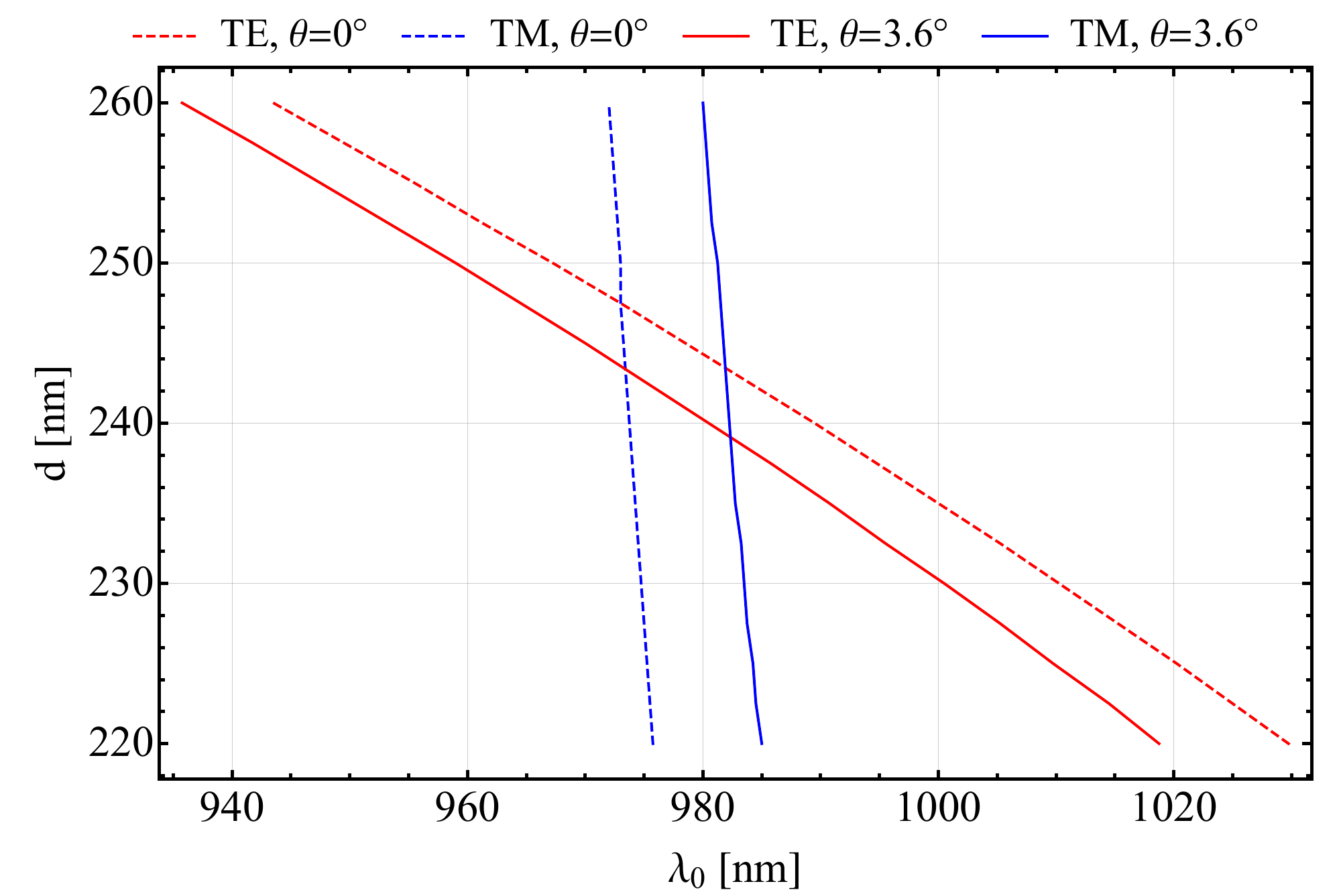}
\caption{Top: RCWA-simulated transmission spectra of the grating with finger depth $d=239$ nm for incident TE-polarized (red) and TM-polarized (blue) light and for $\theta=0$ (dashed) and $\theta=3.6^{\circ}$ (plain) incidence. Bottom: RCWA-simulated positions of the zero-transmission wavelength $\lambda_0$ as a function of the finger depth $d$, for TE-polarized (red) and TM-polarized (blue) light and for $\theta=0$ (dashed) and $\theta=3.6^{\circ}$ (plain) incidence. The other parameters are as stated in the text.}
\label{fig2}
\end{figure}

In order to design the polarization-independent differentiator we consider a one-dimensional grating structure similar to the one previously used to demonstrate first- and second-order spatial differentiation~\cite{Parthenopoulos2021}. As depicted in Fig.~\ref{fig2}, and as a good approximation to the experimental situation discussed in the next Section, the grating is assumed to be periodic in the $x$-direction (period $\Lambda=684$ nm) and consists in SiN trapezoidal fingers with top and bottom finger widths $w_t=366$ nm and $w_b=410$ nm, respectively, and depth $d=249$ nm. The fingers are supported by an underlying SiN slab of thickness $66$ nm, the total thickness of the unpatterned film being $t=315$ nm. The refractive index in the wavelength range of interest is taken to be constant and equal to 2.14 (see Sec.~\ref{sec:exp}). 

The transmission properties of the grating under monochromatic illumination with linearly polarized plane waves are simulated by Rigorous Coupled Wave Analysis (RCWA) using MIST~\cite{MIST} by discretizing the structure in 20 layers and using a 25 mode expansion. The RCWA-simulated spectra of the grating for incident TE-polarized ($\phi=90^{\circ}$) and TM-polarized ($\phi=0^{\circ}$) light impinging on the grating at normal ($\theta=0^{\circ}$) and oblique ($\theta=3.6^{\circ}$) incidence are shown in Fig.~\ref{fig2}. At normal incidence, zero-transmission (high-reflectivity) guided-mode resonances for TM- and TE-polarized light are observed at (680, 972) nm and (755, 992) nm, respectively. Even tough the grating thickness is substantially larger than the underlying slab thickness, these resonances can be respectively attributed to excitation of leaky TM$_{1}$, TM$_{0}$, TE$_{1}$ and TE$_{0}$ waveguide modes (see Supplementary Material). We will focus in the following on the TM$_{0}$ and TE$_{0}$ resonances.

At an incidence angle of 3.6$^{\circ}$, these two resonances are observed to occur simultaneously at $\sim 982.5$ nm. At this oblique incidence, other resonances also appear due to the excitation of guided modes corresponding to odd-symmetry superpositions of contrapropagating waves into the grating. It is indeed well-known that, at normal incidence and for a perfect, infinite grating structure, only even-symmetry superpositions--with respect to the plane of symmetry of the grating--can be excited by plane waves~\cite{Magnusson1993,Rosenblatt1997,Fan2003}. At oblique incidence, however, the odd-symmetry superpositions can also be excited, yielding additional resonances in the transmission spectrum. These even- and odd-symmetry superposition resonances are shifted with the incidence angle in a direction that depends on the guided mode dispersion close to the resonance considered. Here, the TM$_{0}$ and TE$_{0}$ resonances considered are respectively red- and blue-shifted. Since the normal incidence TM$_{0}$ resonance occurs at a slightly lower wavelength than the normal incidence TE$_{0}$ resonance, it is then possible to have both resonances overlap for a small angle of incidence, of $\sim3.6^{\circ}$ here.

The parameters of the structured films can thus be adjusted so as to achieve polarization-independent resonance at a given operation wavelength and incidence angle. To illustrate the dependence of these resonances with the design parameters, Fig.~\ref{fig2} shows as an example the position of the zero-transmission resonance as a function of the finger depth $d$ for both normal or oblique incidence and for both TE and TM polarizations. For the parameters chosen here, the TE-polarized light resonances show a much stronger dependence on the finger depth than those for TM-polarized light. While we focus in the following on the polarization-independent resonance achieved at an oblique incidence of $3.6^{\circ}$, let us remark that polarization-independent resonance at normal incidence could also be achieved by a slight increase ($\sim 8.5$ nm) of the finger depth. In this way, second-order spatial differentiation, as demonstrated in~\cite{Parthenopoulos2021}, could be realized, albeit in a polarization-independent fashion (see Supplementary Material).

\subsection{First-order spatial differentiation at oblique incidence}\label{sec:fem}

As abovementioned, first-order spatial differentiation of the transverse profile of optical beams using guided-mode resonance gratings at oblique incidence was suggested in~\cite{Golovastikov2014} and recently demonstrated with various grating structures and materials~\cite{Bykov2018,Dong2018,Yang2020,Parthenopoulos2021}. The mechanism of spatial differentiation can be simply understood on the basis of a one-dimensional spectral decomposition of the incoming beam and considering the transfer function of the grating in Fourier space based on a simple coupled-mode model~\cite{Bykov2015}. Denoting by $G_\textrm{inc}(k_x)$ and $G_\textrm{tr}(k_x)$ the spatial angular spectra of the incident and transmitted fields, respectively, the spatial transfer function of the grating is given by~\cite{Golovastikov2014}
\begin{equation}
H(k_x)\equiv G_\textrm{tr}(k_x)/G_\textrm{inc}(k_x),
\end{equation}
where $k_x$ is the wavevector component of the plane wave impinging with an incident angle $\alpha=\theta+\arcsin(k_x/k_0)$ with respect to the $z$-axis and where $k_0=2\pi/\lambda_0$ is the resonance wavevector. For a weakly focused incident beam with a characteristic angular distribution width $\Delta$ smaller than $\theta$ and in the vicinity of the resonance, the transfer function can be shown to be approximately given by
\begin{equation}
H(k_x)\simeq iCk_x,
\end{equation}
where $C$ is a constant. This transfer function, leading to a first-order differentiation in real space, is expected to be valid at least when the transmission spectrum of the grating considered can be well-reproduced by the two-mode model of~\cite{Bykov2015} (see e.g.~\cite{Parthenopoulos2021} for a detailed comparison of the transmission spectra of similar gratings under various incidence conditions). Let us also remark that finite-size and collimation effects can also affect the position, width and shape of the observed resonances~\cite{Magnusson1993,Loktev1997,Glasberg1998,Jacob2000,Bendickson2001,Peters2004,Niederer2004}. We refer to the reader to Ref.~\cite{Toftvandborg2021} for a detailed study of these effects with similar, albeit thinner, gratings structures.

\begin{figure}[h]
\includegraphics[width=\columnwidth]{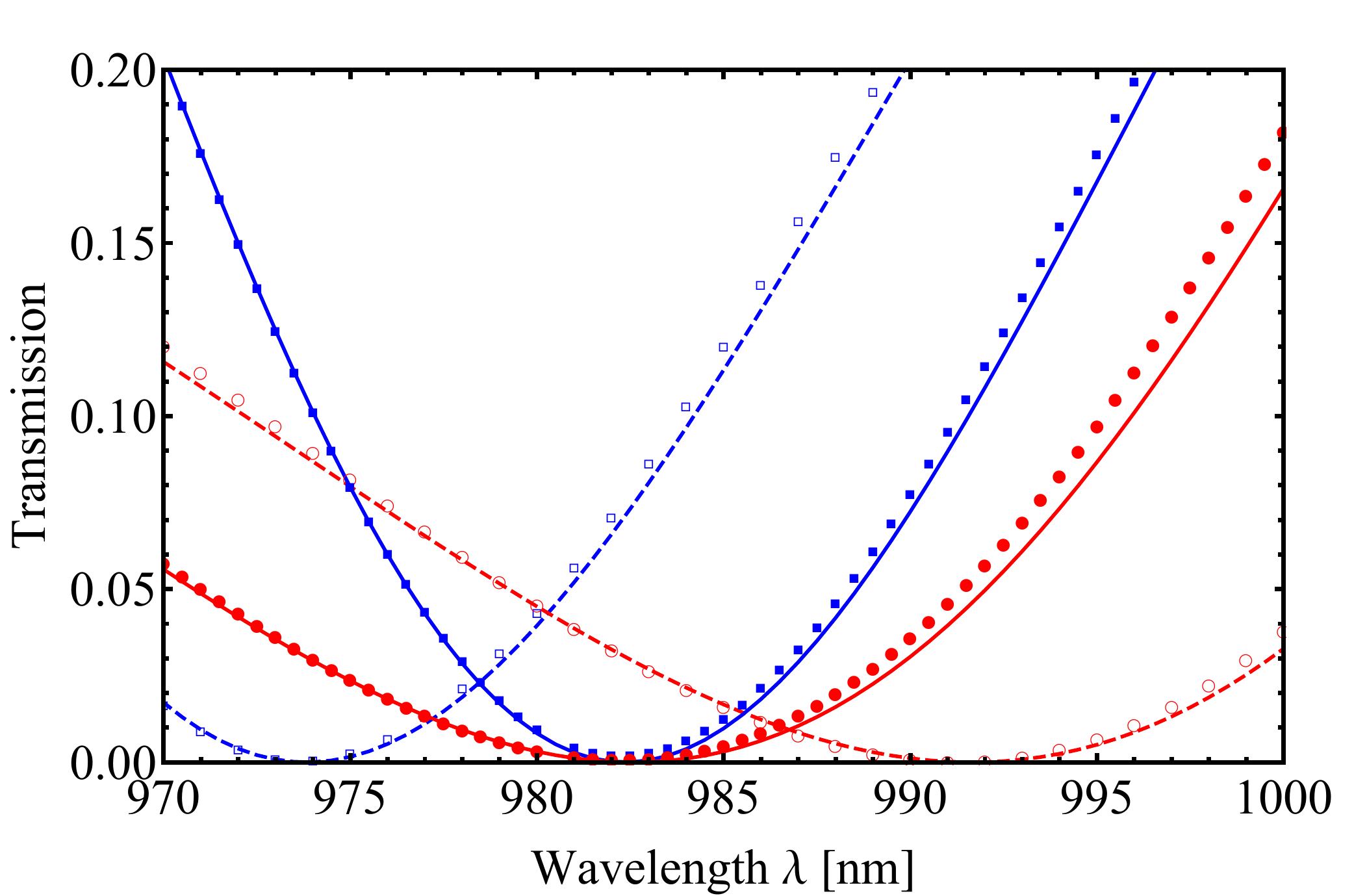}
\caption{Comparison between the RCWA-simulated (lines) and FEM-simulated (symbols) spectra at normal (dashed) and $3.6^\circ$ oblique (plain) incidence for TE (red) and TM (blue) incident polarizations.}
\label{fig4}
\end{figure}


To check these predictions in the geometry of interest FEM simulations  were performed using Comsol Multiphysics by considering a 2D model of the trapezoidal grating, periodic in the $x$-direction with an extension of 200 $\mu$m and infinite in the $y$-direction. To match the experimental situation the Gaussian beam is assumed to have a waist (radius) of $w_0=40$ $\mu$m, and thereby an angular spectrum width $\Delta=\lambda/\pi w_0\simeq 0.45^\circ$. Figure~\ref{fig4} presents a comparison the RCWA-simulated and FEM-simulated spectra at normal and oblique incidence for both TE and TM polarizations. The good matching between the plane-wave and the Gaussian simulations shows that finite-size and collimation effects are small for these parameters. 

\begin{figure}[h]
\includegraphics[width=\columnwidth]{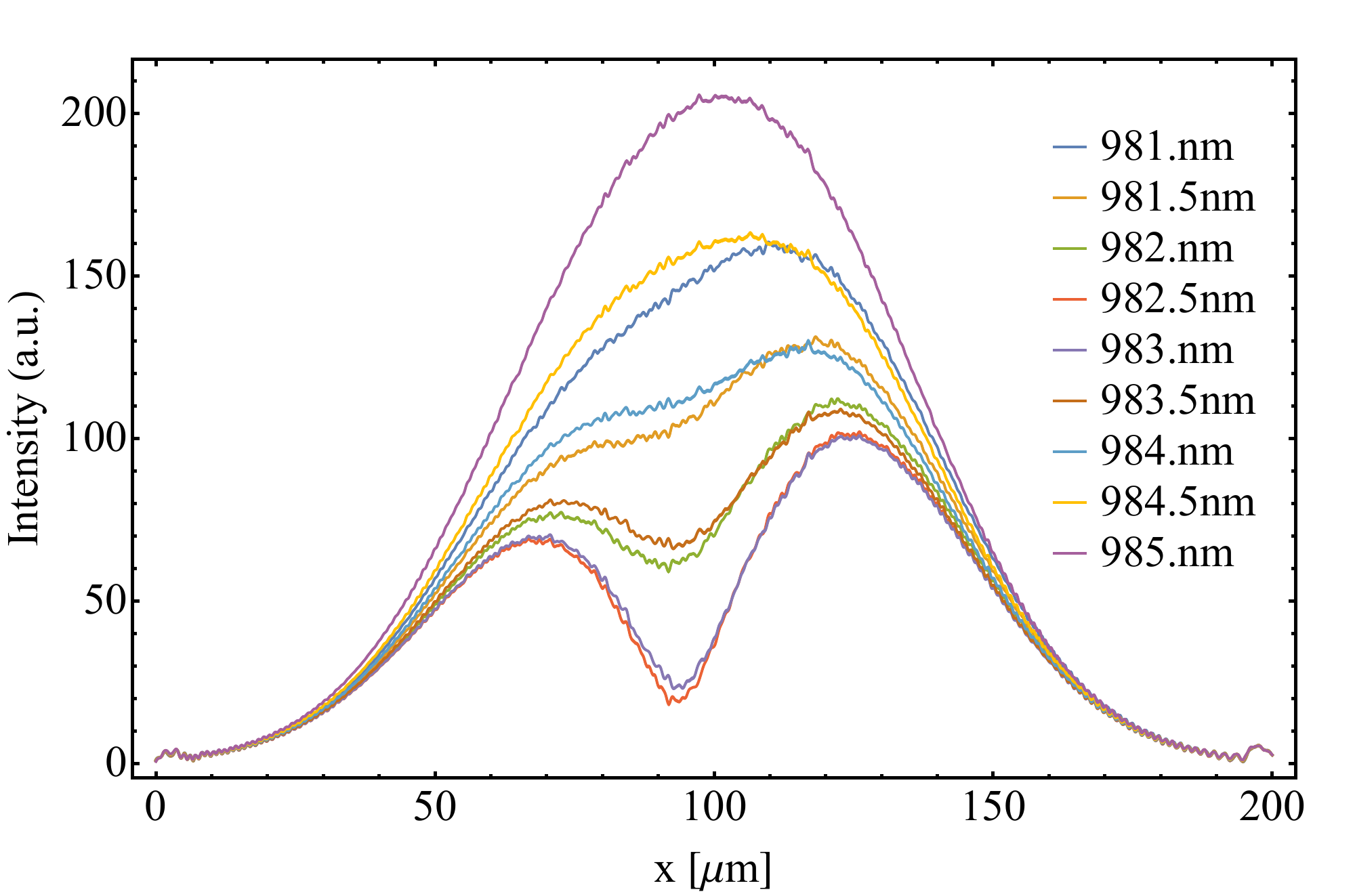}\\
\includegraphics[width=\columnwidth]{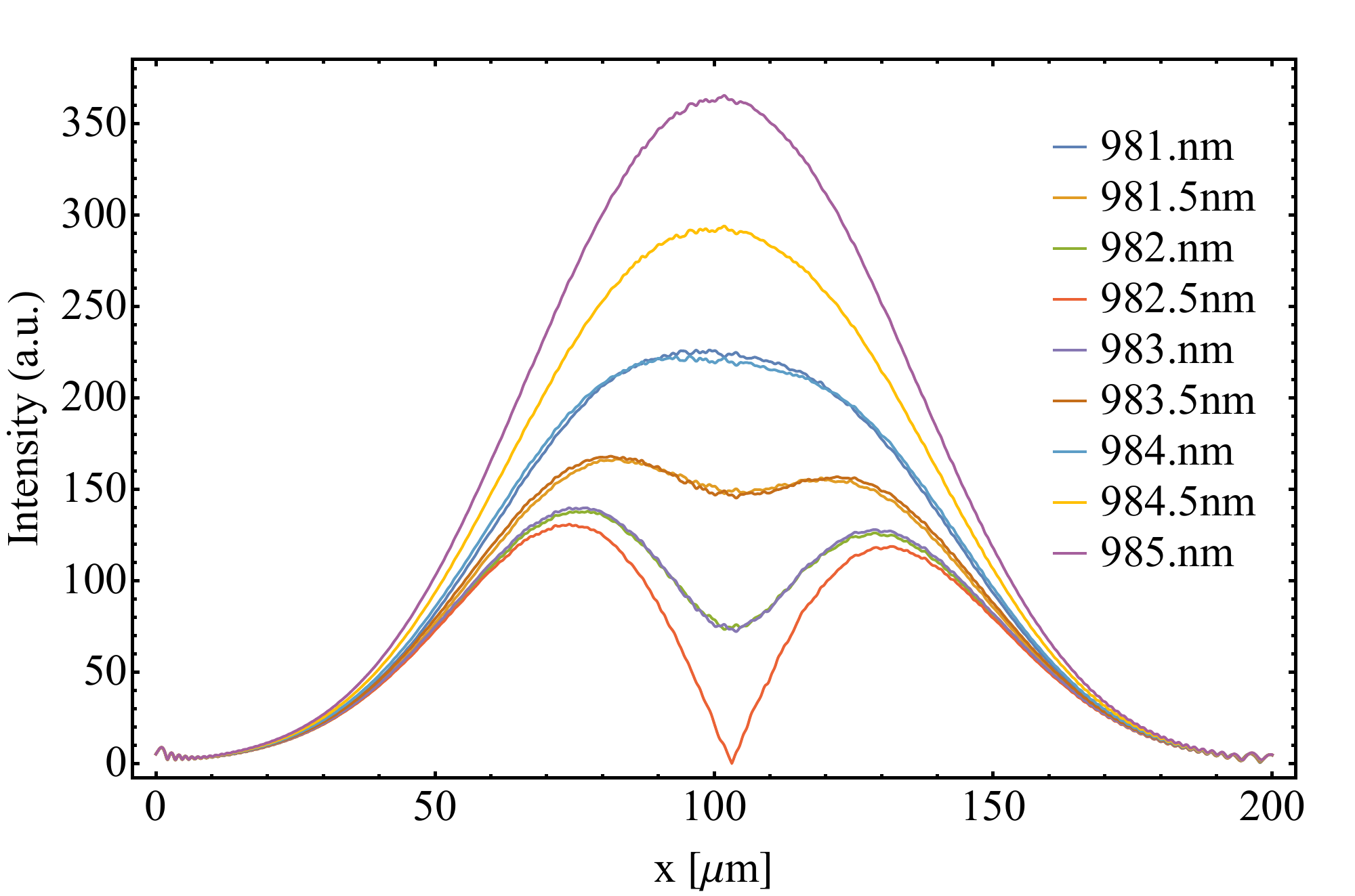}
\caption{FEM-simulated transverse intensity profiles of the transmitted TE-polarized (top) and TM-polarized (bottom) beams for an incidence angle $\theta=3.6^{\circ}$ and various wavelengths close to the common TE$_0$/TM$_0$ resonance.}
\label{fig42}
\end{figure}

Last, Fig.~\ref{fig42} shows the transverse intensity profiles of the TE- and TM-polarized transmitted beams at a $3.6^\circ$ oblique incidence for wavelengths close to the common resonant wavelength of 982.5 nm. Profiles consistent with first-order spatially differentiated Gaussian beams are clearly observed close to resonance for both polarizations. A slight asymmetry in the spatial differentiation of the TE-polarized beam can be also observed. We attribute this to the close proximity of the odd-superposition TE$_{1}$ resonance at 820 nm, whose wavelength separation from the 982.5 nm resonance is comparable with the odd superposition TE$_{0}$ resonance at 1125nm (see Fig.~\ref{fig2}). Indeed, the spatial differentiation predicted by the coupled two-mode model of Ref.~\cite{Golovastikov2014} relies on a transfer function based on a well-isolated double-Fano resonance~\cite{Bykov2015}. However, the presence of an additional nearby resonance can be expected to alter this picture and thereby affect the differentiation operation (see Supplementary Material).


\section{Experimental methods and results}\label{sec:exp}

\begin{figure}[h]
\includegraphics[width=\columnwidth]{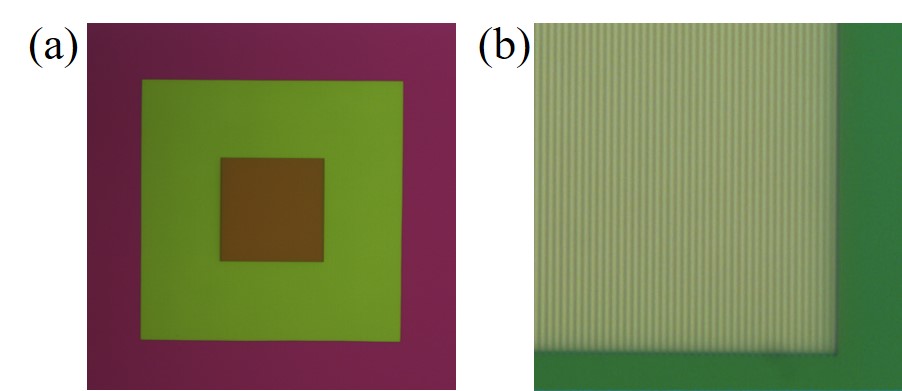}\\\vspace{0.2cm}
\includegraphics[width=\columnwidth]{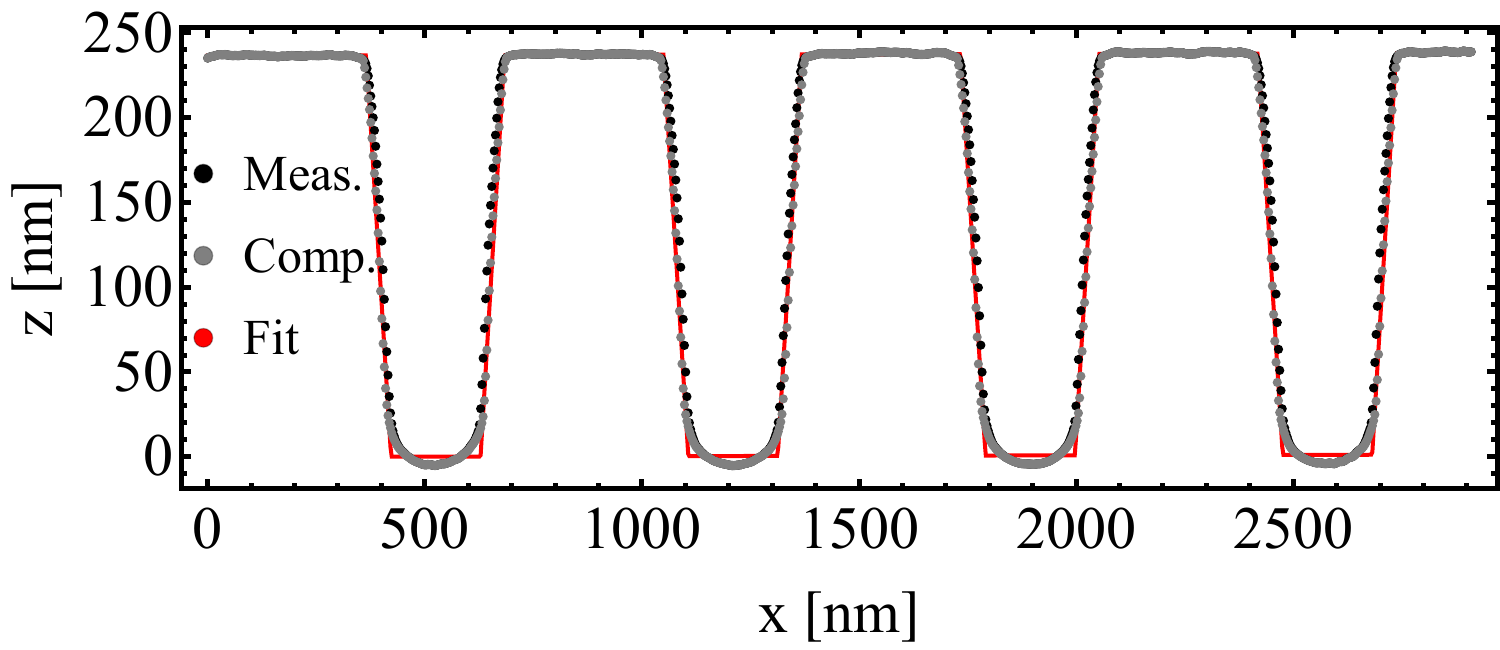}\\
\includegraphics[width=\columnwidth]{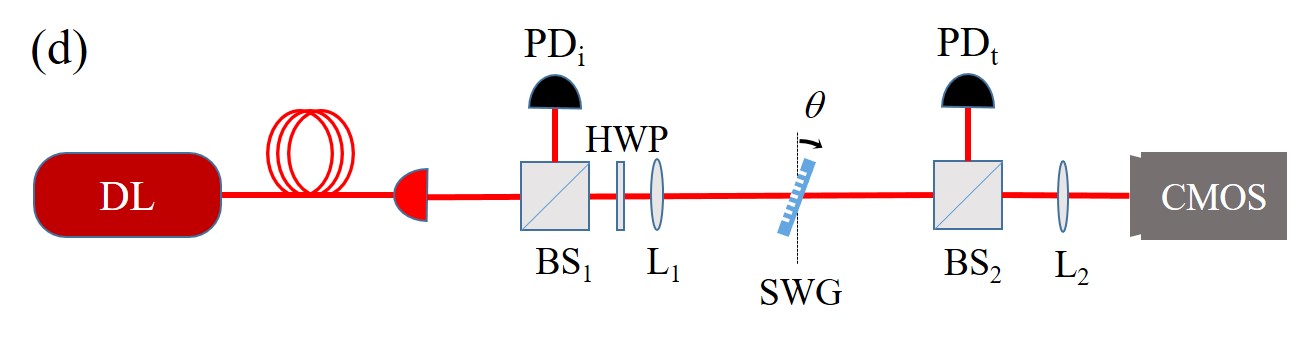}
\caption{(a) Topview picture of the 200 $\mu$m-square grating patterned on a 500 $\mu$m-square SiN thin film (green) suspended on a silicon frame (black). (b) Zoom on one corner of the grating. (c) Result of AFM scan profilometry. The black points show the measured height, the grey points show the height after compensation from the AFM tip curvature and the red line shows the results of a trapezoidal fit to the grey data points. (d) Experimental setup for the optical characterization (see text for details). DL: diode laser, BS$_1$ and BS$_2$: 50/50 beamsplitters, PD$_\textrm{i}$ and PD$_\textrm{t}$: photodectectors, HWP: halfwave plate, L$_1$ and L$_2$: lenses, SWG: subwavelength grating, CMOS: camera.}
\label{fig5}
\end{figure}

The sample used in this work was fabricated by Norcada, Inc.~\cite{Norcada} and its topology characterized using the AFM profilometry method of Ref.~\cite{Darki2021}. The refractive index and total thickness of the unpatterned film were measured by ellipsometry~\cite{Nair2017}. Figures~\ref{fig5}(a) and (b) show images of the 200 $\mu$m-square grating patterned on a 500 $\mu$m-square suspended, low stress SiN film and Fig.~\ref{fig5}(c) shows the result of an AFM scan of one part of the patterned area. The observed profile is close to trapezoidal, as observed with similar structures fabricated using Electron Beam Lithography and chemical etching~\cite{Nair2019,Parthenopoulos2021,Darki2021,Toftvandborg2021}, the main difference being the increased film thickness ($\sim 315$ nm versus 200 nm) in order to achieve polarization-independent resonances in the wavelength range of 915-985 nm of the laser available for the sample characterization. Another slight difference is a more pronounced curvature at the bottom of the grooves. For simplicity and ease of comparison with our previous structures we still make use, though, of a fit with a trapezoidal profile for the AFM scan analysis and the simulations. 

The geometrical parameters of the gratings are extracted from the results AFM scans in various parts of the grating and after correction from the AFM tip radius curvature, as described in~\cite{Darki2021}. Since the AFM tip angle is larger than the grating finger wall angle, this latter is adjusted in the simulations to match the observed transmission spectra at normal incidence, as discussed in~\cite{Darki2021}.

\begin{figure}[h]
\includegraphics[width=\columnwidth]{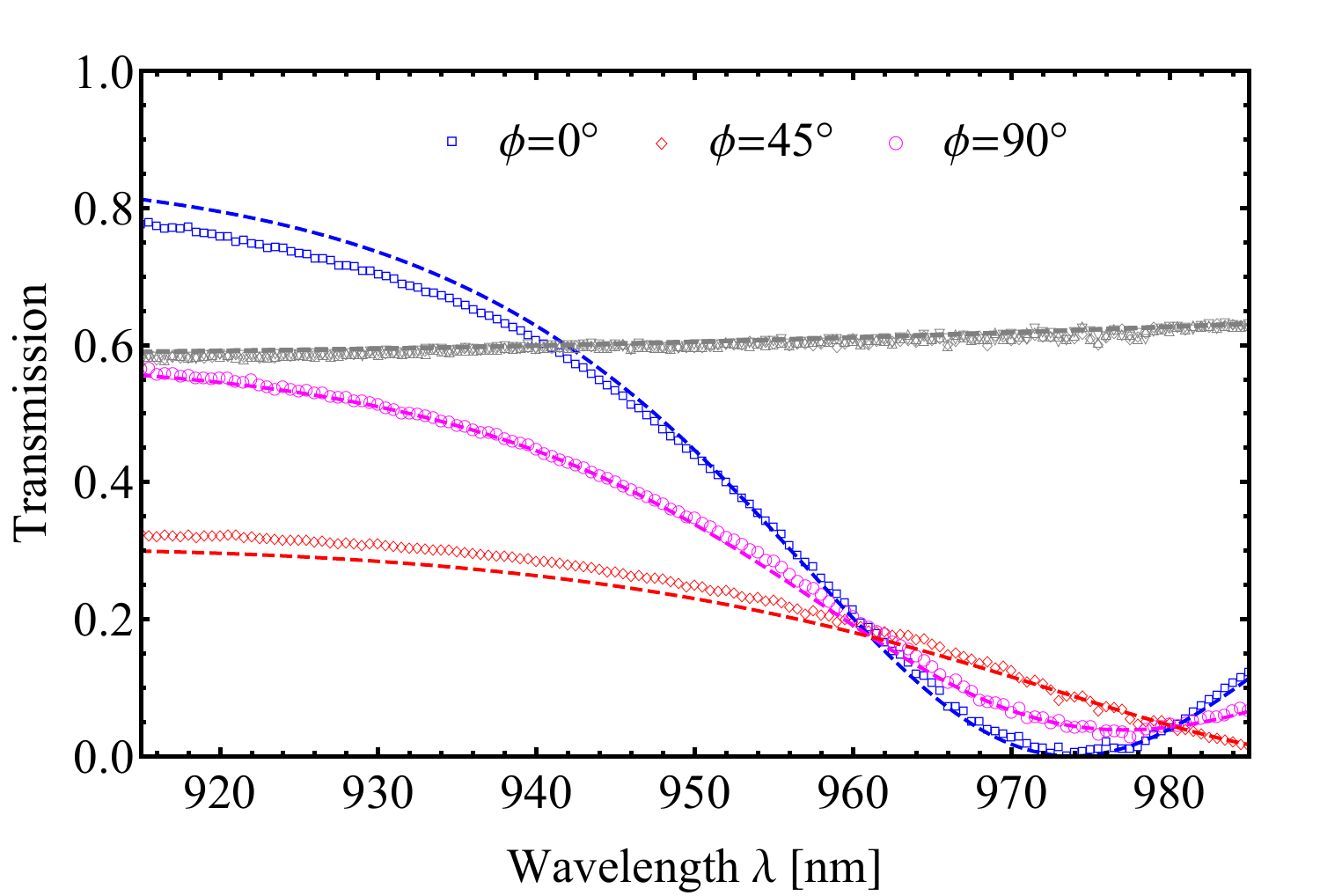}\\
\includegraphics[width=\columnwidth]{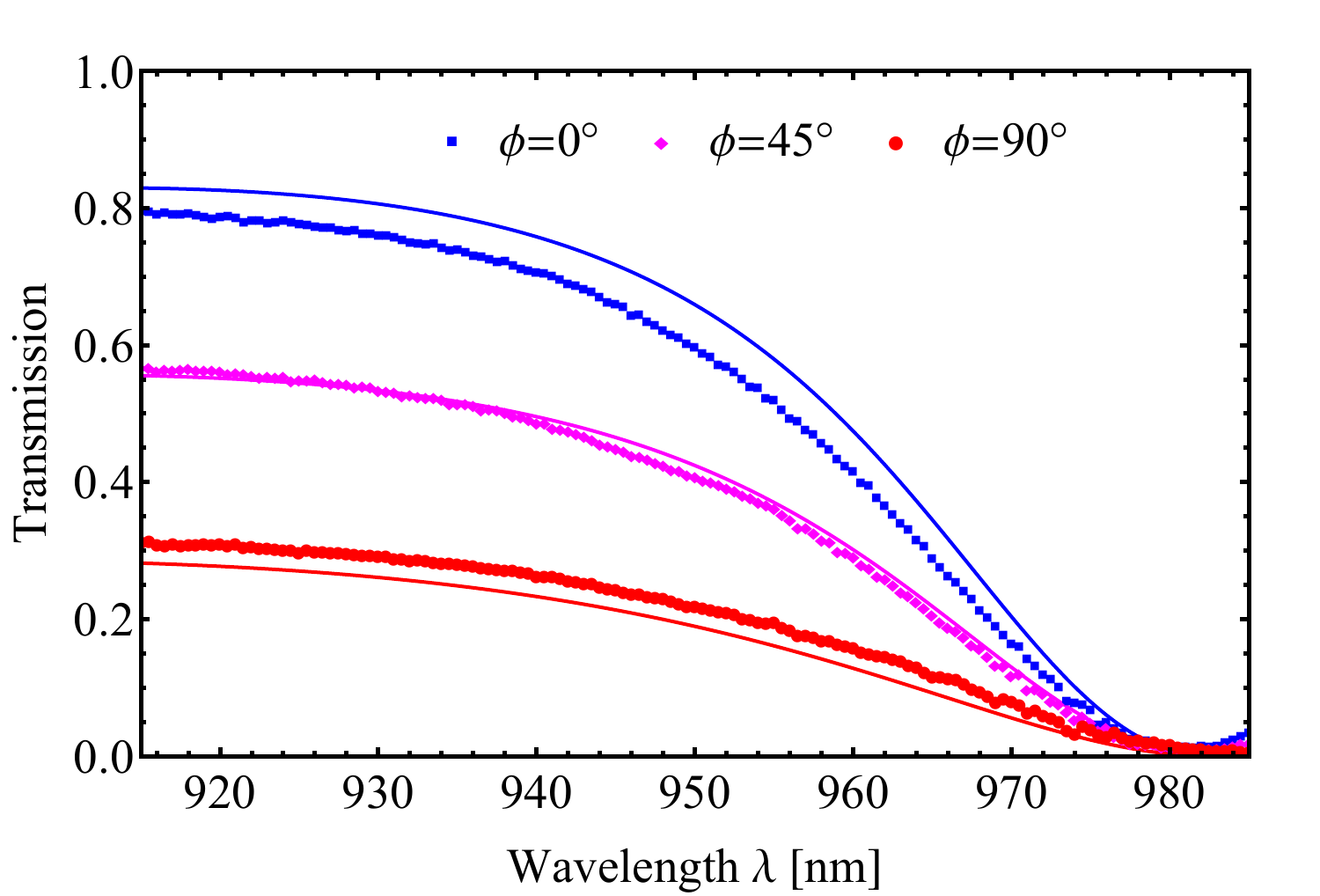}
\caption{Experimental transmission spectra at normal (top, $\theta=0^\circ$) and oblique (bottom, $\theta=3.6^{\circ}$) incidence for $\phi=0^{\circ}$, $\phi=45^{\circ}$ and $\phi=90^{\circ}$-polarized light. The dashed lines show the corresponding RCWA simulated spectra. The grey data points and line and show the case when the beam impinges on the non-patterned suspended dielectric slab. }
\label{fig6}
\end{figure}

The transmission spectrum of the grating is measured using the experimental setup described in~\cite{Parthenopoulos2021}. In brief, the light from a monochromatic, tunable external cavity diode laser (Toptica DLC Pro 915-98 nm) is coupled into a single-mode, polarization-maintaining fiber and subsequently focused onto the grating at normal or oblique incidence using an achromatic 75 mm-focal length doublet (L$_1$). The polarization of the incident light is set by an achromatic half-wave plate (HWP). Half of the transmitted light is collected on a photodetector (PD$_t$) and referenced to the input power monitored by another photodetector before the sample (PD$_i$). The light transmitted by beamsplitter BS$_2$ is collected by a 75 mm-focal length and a CMOS camera in order to provide an image of the transverse profile of the transmitted beam at the grating position with a magnification of approximately 3. The transmission of the sample under various incidence angles and incident polarizations is obtained by scanning the laser wavelength and normalizing the referenced transmitted photodetector signal by that without sample~\cite{Nair2017,Nair2019}. As discussed in Sec.~\ref{sec:design} and in Ref.~\cite{Toftvandborg2021}, the large area of the grating allows for operating with a waist (radius) of $w_0=40$ $\mu$m, thus minimizing finite-size and collimation effects and improving the quality of the spatial differentiation.

The normalized transmission of the grating at normal incidence ($\theta=0^{\circ}$) are shown in Fig.~\ref{fig6} for $\phi=0^\circ$, $\phi=45^\circ$ and $\phi=90^\circ$--linearly polarized light, together with the corresponding RCWA-simulated spectra. As a reference, the spectra obtained by focusing the beam on the suspended dielectric slab outside the patterned area are also shown and observed to be polarization-independent and consistent with the thickness of 315 nm and refractive index of 2.14 independently determined by ellipsometry (gray dashed line). There is an overall good agreement between the simulated and experimental spectra, the slight discrepancies being most likely due to our approximate modelling of the grating profile, or possibly small structural inhomogeneities.

\begin{figure}[h]
\includegraphics[width=\columnwidth]{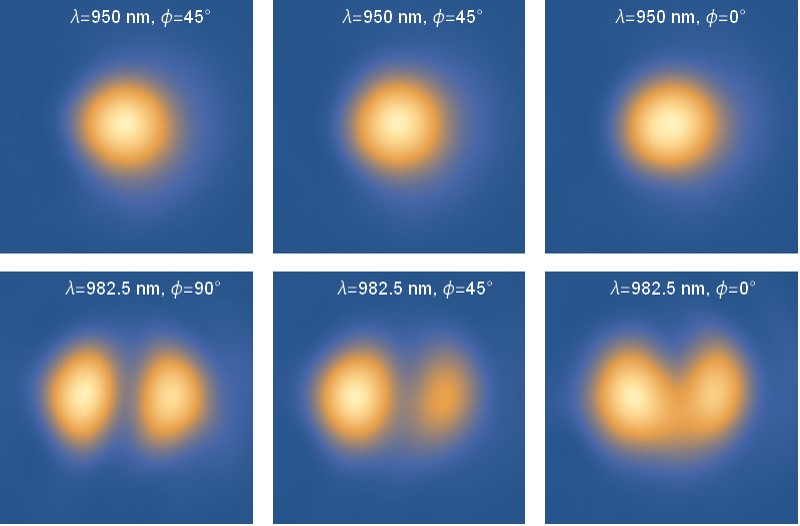}
\caption{Transverse intensity profiles of the transmitted beam at oblique incidence ($\theta=3.6^\circ$) for $\phi=0^{\circ}$, $45^{\circ}$ and $90^{\circ}$--polarized light at an off-resonant wavelength of 950 nm (top) and at the resonant wavelength of 982.5 nm (bottom).}
\label{fig7}
\end{figure}

Figure~\ref{fig6} also shows the experimental spectra for the same incident polarizations, but at a $3.6^\circ$ oblique incidence, as well as the corresponding RCWA-simulated spectra. The resonances for the 3 polarizations are observed to be close to degenerate and with an equal transmission of $\sim 1$ \% at 982.5 nm. Figure~\ref{fig7} shows the measured transverse intensity profiles of the transmitted beam at oblique incidence and at resonance for the 3 polarizations, as well as the corresponding profiles away from resonance for comparison. Fits of the measured profiles away from resonance with a Gaussian intensity profile ($\sim\exp[-2(x^2+y^2)/w_0^2]$) yield a $\chi^2$ of 0.99, while fits of the measured profiles at resonance with a first-order differentiated Gaussian beam intensity profile ($\sim x^2\exp[-2(x^2+y^2)/w_0^2]$) yield a $\chi^2$ of 0.97, 0.88 and 0.88 for the TM, 45$^\circ$ and TE polarizations. As discussed in Sec.~\ref{sec:fem} and in the Supplementary Material, we attribute the comparatively lower quality of the spatial differentiation for TE-polarized light than for TM-polarized light to the presence of the nearby TE$_1$ guided mode resonance. In the future such an effect could be minimized by inverse design using e.g. topology optimization~\cite{Christiansen2021} of the grating parameters in order to increase the separation between these resonances.

\section{Conclusion}\label{sec:conclusion}

A one-dimensional, symmetric subwavelength grating patterned on a thin, suspended silicon nitride film was designed in order to achieve a polarization-independent guided-mode resonance at a small incidence angle. FEM simulations predicted polarization-independent first-order spatial differentiation to occur at this resonance under oblique incidence. These predictions were verified experimentally by measurements of the transmission spectrum of the grating under illumination with light with various polarizations and incidence angles, as well as measurements of the transverse intensity profiles of the transmitted beam. 

Routes to improve the quality of the spatial differentiation by improving the spectral isolation of the resonances or to achieve polarization-independent second-order spatial differentiation by operating at normal incidence were discussed as well. Let us further note that piezoelectric tuning of the optical response as demonstrated in Ref.~\cite{Nair2019} could also be applied in order to extend the tuning range of the spatial differentiation.



\section*{Funding}
Independent Research Fund Denmark.

\section*{Acknowledgments}
We are grateful to Norcada for their assistance with the design and fabrication of the samples used in this study.

\section*{Disclosures} 
The authors declare no conflicts of interest.


\title{Polarization-independent optical spatial differentiation\\with a doubly-resonant one-dimensional guided-mode grating\\ (Supplementary Material)}

\author{Ali Akbar Darki$^1$, S{\o}ren Peder Madsen$^2$ and Aur\'{e}lien Dantan$^1$}\email{dantan@phys.au.dk}

\address{$^1$Department of Physics and Astronomy, Aarhus University, DK-8000 Aarhus C, Denmark\\
$^2$Department of Mechanical and Production Engineering, Aarhus University, DK-8000, Aarhus C, Denmark}

\date{\today}

\maketitle


\section{Guided-mode resonance field distributions}

\begin{figure}[h]
\includegraphics[width=0.35\textwidth]{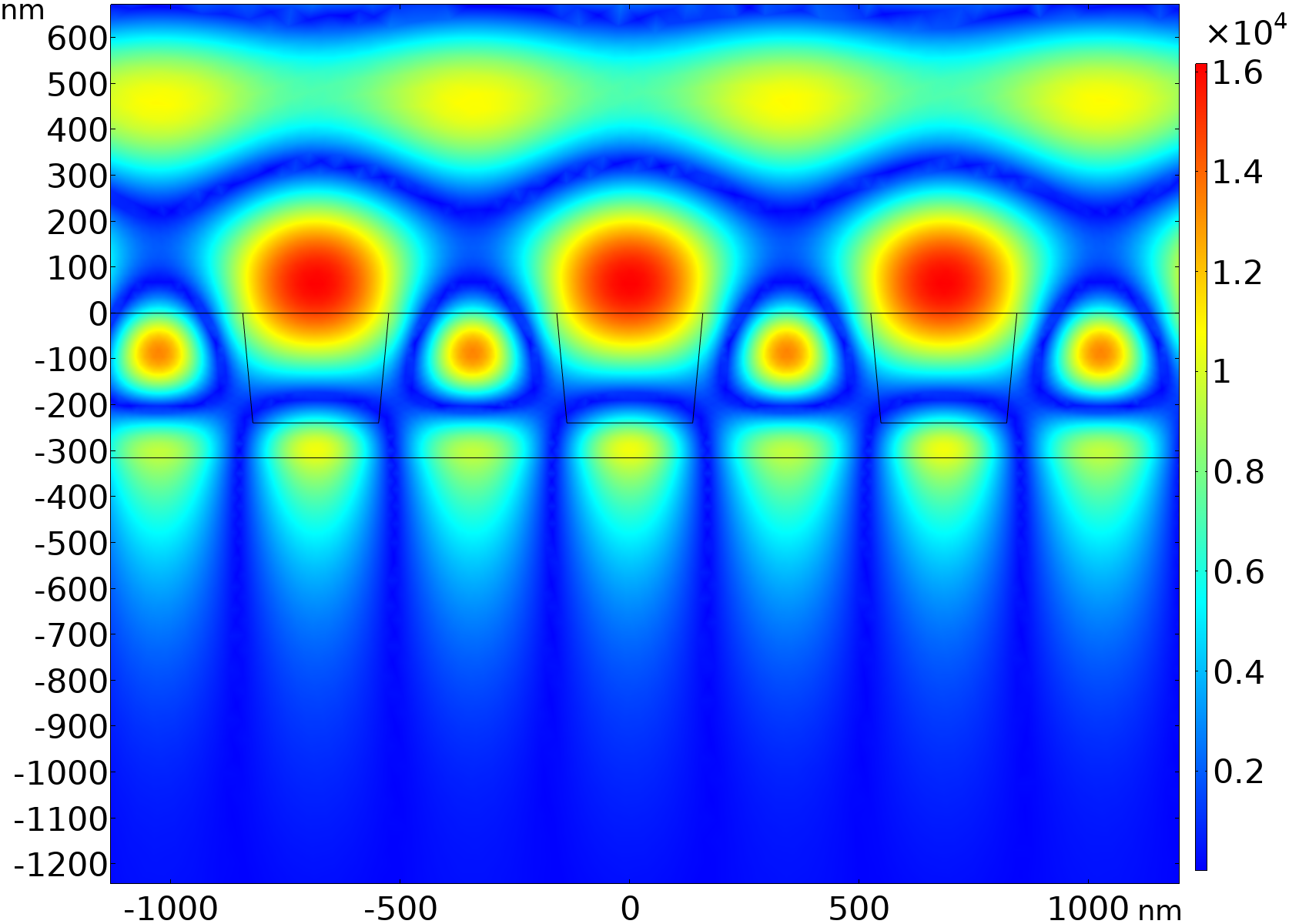}
\includegraphics[width=0.35\textwidth]{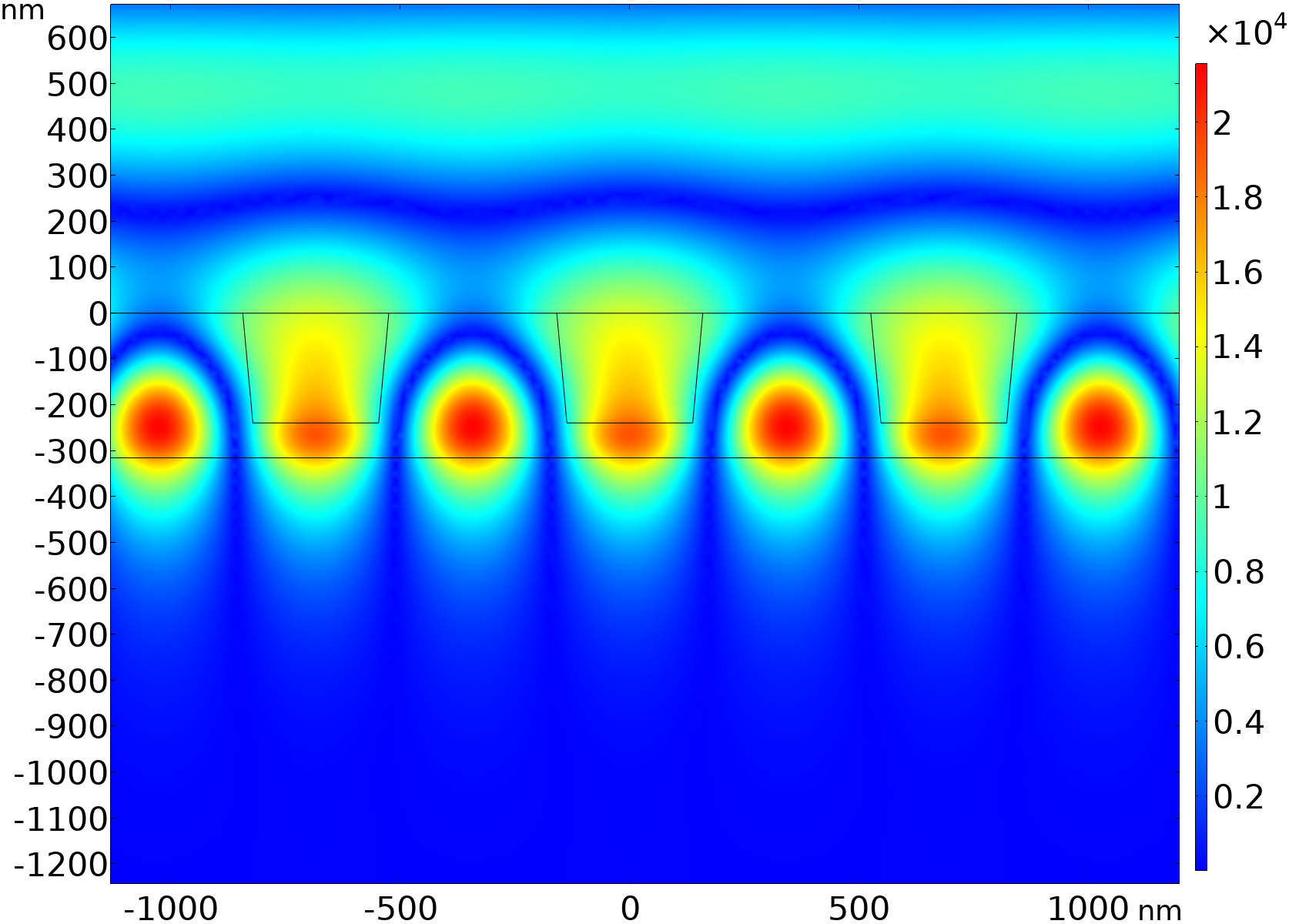}\\
\includegraphics[width=0.35\textwidth]{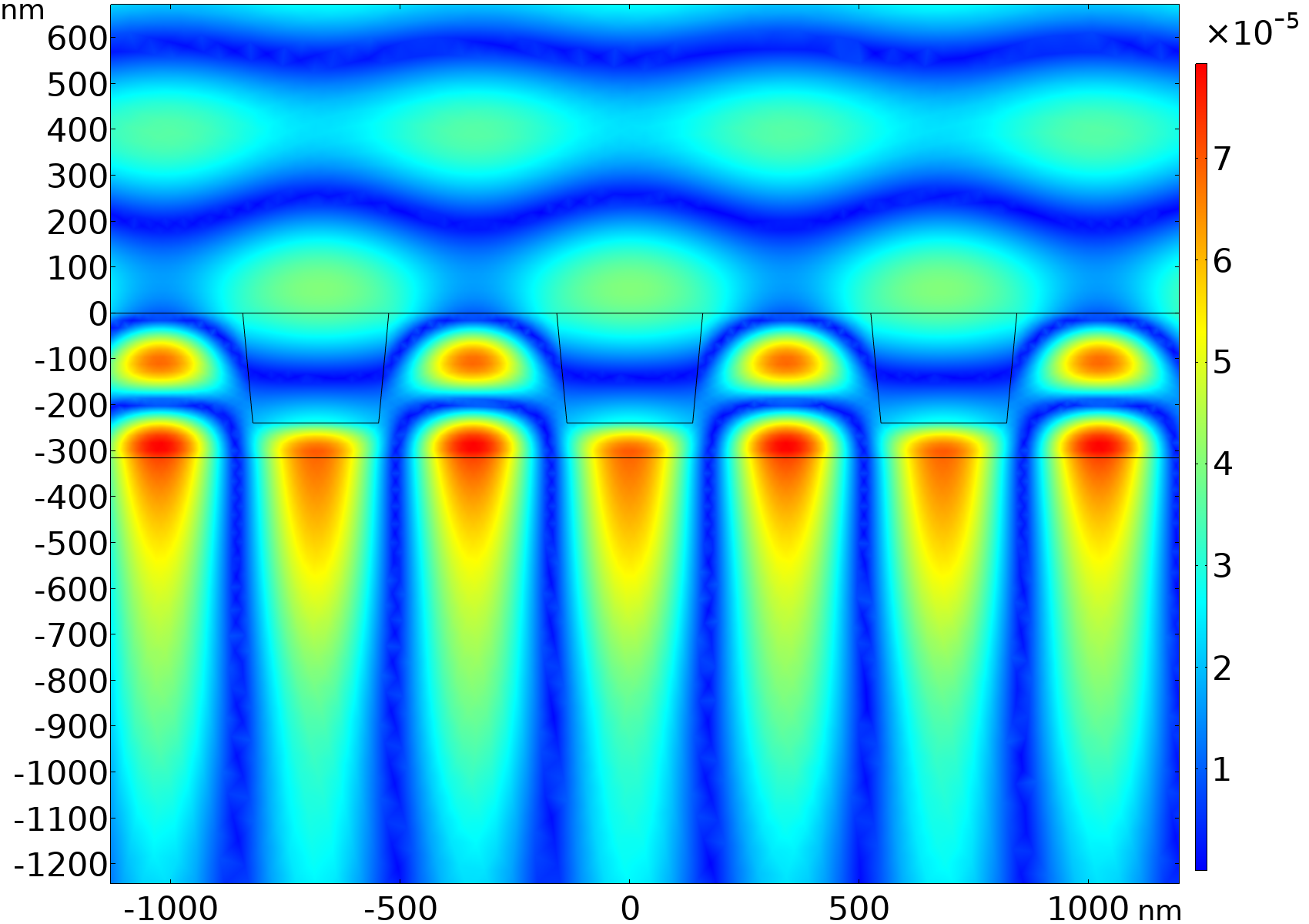}
\includegraphics[width=0.35\textwidth]{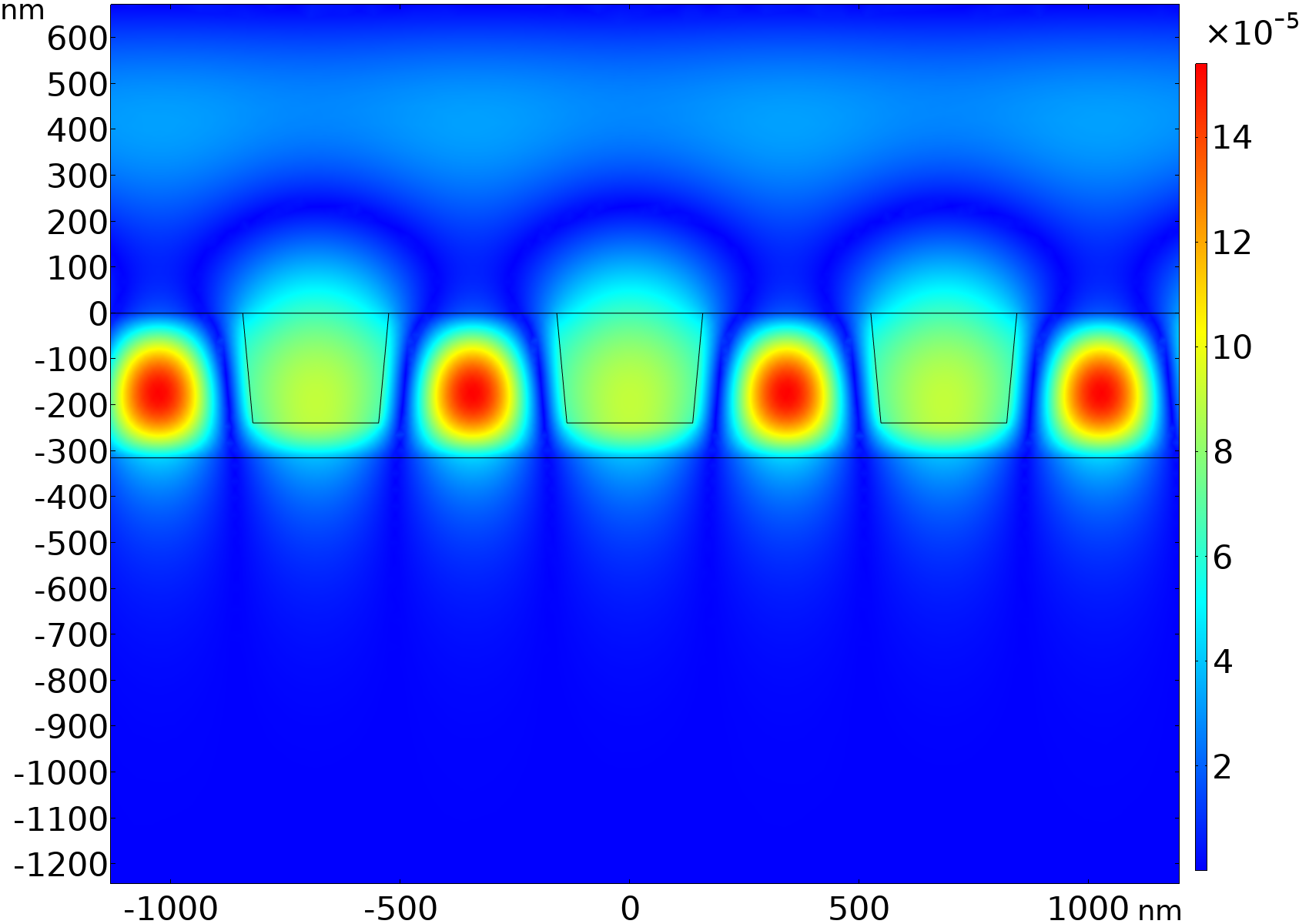}
\caption{FEM-simulated electric/magnetic (top/bottom) field intensity distributions in the vicinity of the grating investigated in Sec.~II of the main text  for TE-polarized light (top) at 745 nm (left) and 992 nm (right), and for TM-polarized light (bottom) at 689 nm (left) and 974 nm (right).}
\label{fig:fields}
\end{figure}

Figure~\ref{fig:fields} shows the FEM-simulated electric/magnetic (left/right) field intensity distributions in the vicinity of the grating investigated in Sec.~II of the main text, for TE-polarized light (top) at 745 nm (left) and 992 nm (right), and for TM-polarized light (bottom) at 689 nm (left) and 974 nm (right) impinging on the grating at normal incidence. These wavelengths can thus be respectively identified as leaky TE$_1$, TE$_0$, TM$_1$ and TM$_0$ guided-mode resonances.

\section{Second-order spatial differentiation simulations}

\begin{figure}[h]
\includegraphics[width=0.49\textwidth]{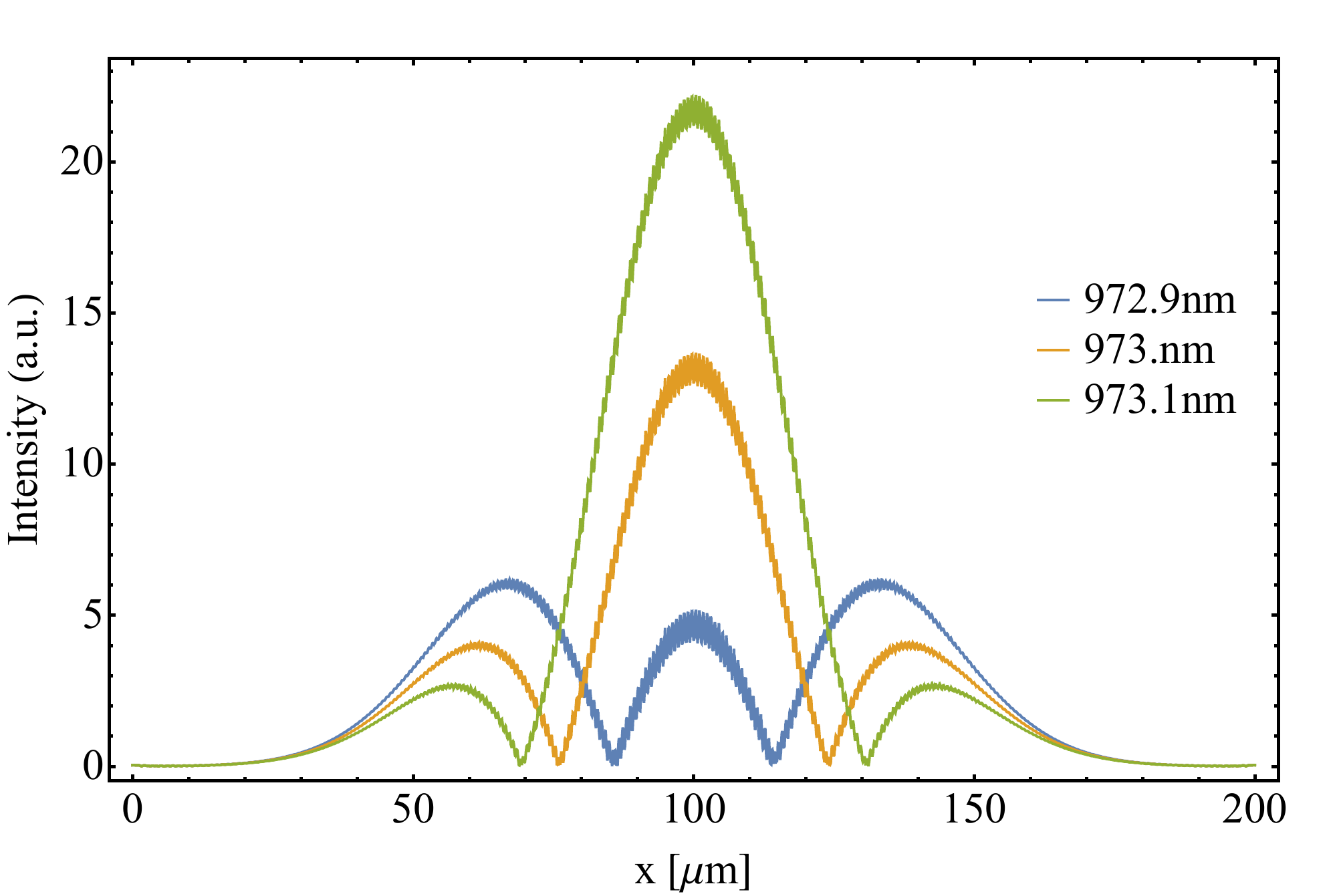}
\includegraphics[width=0.49\textwidth]{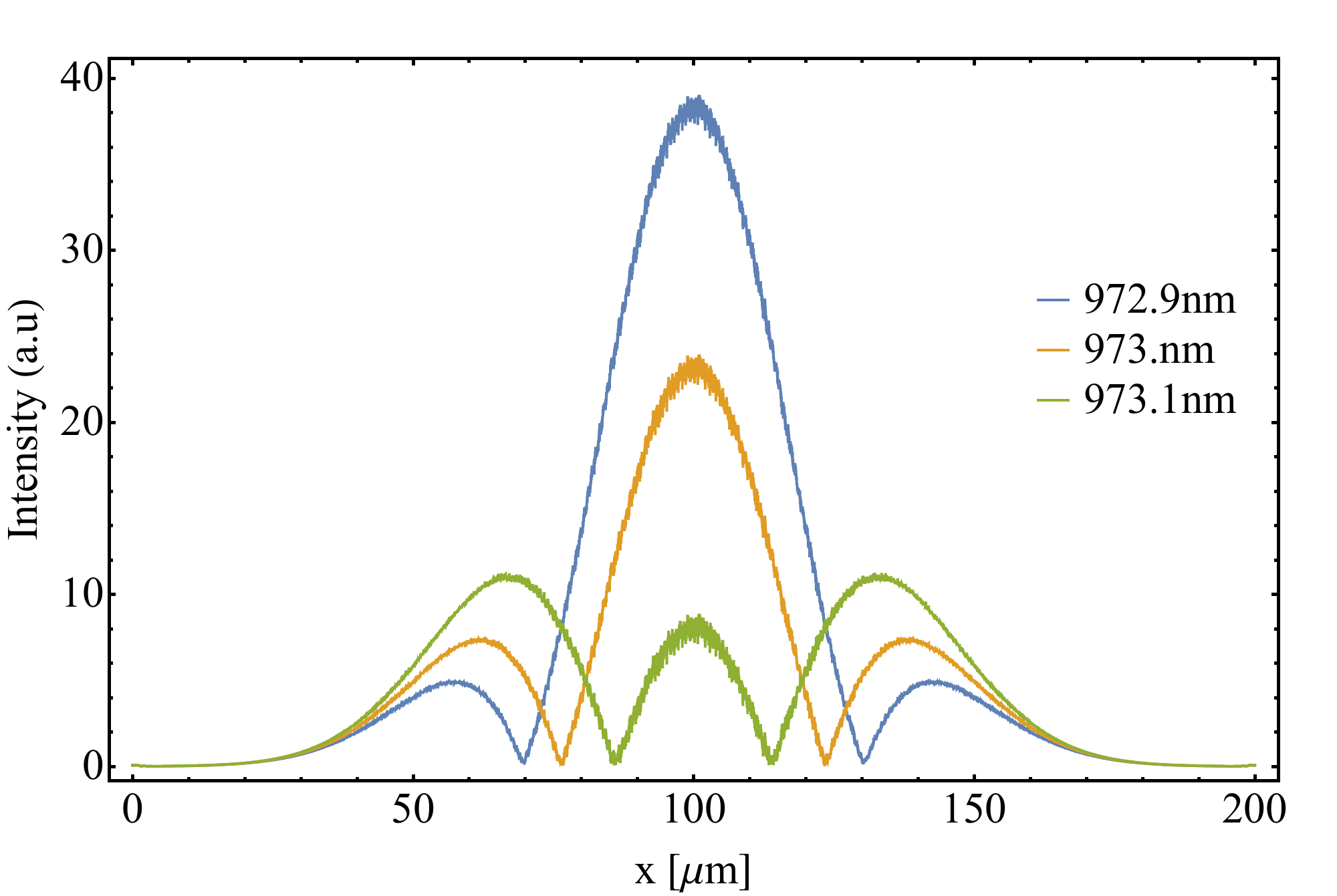}
\caption{FEM-simulated transmitted beam intensity profiles of TE-polarized (left) and TM-polarized (right) light impinging at normal incidence on a 247.47 nm-deep grating for wavelengths close to the common TE$_0$/TM$_0$ resonance.}
\label{fig:d2475}
\end{figure}

To investigate the possibility of performing polarization-independent second-order differentiation, we simulated the transmitted beam profile for TE- and TM-polarized light impinging at normal incidence on a grating with a finger depth of $d=247.47$ nm, the other parameters being the same as those of Sec.~II of the main text. The results are shown in Fig.~\ref{fig:d2475}, where profiles consistent with second-order spatial differentiation of a Gaussian beam with waist 30 $\mu$m are observed around the common TE$_0$/TM$_0$ normal incidence resonance at 973 nm. Let us remark that, for the grating parameters used here, achieving good quality polarization-independent second-order spatial differentiation at normal incidence requires a rather fine tuning of the finger depth, because of the rather strong sensitivity of the TE$_0$ resonance position with respect to $d$. We expect, though, that this sensitivity could be reduced by a more systematic optimization of the grating parameters. 

\section{Effect of higher-order resonance}

\begin{figure}[h]
\includegraphics[width=0.49\textwidth]{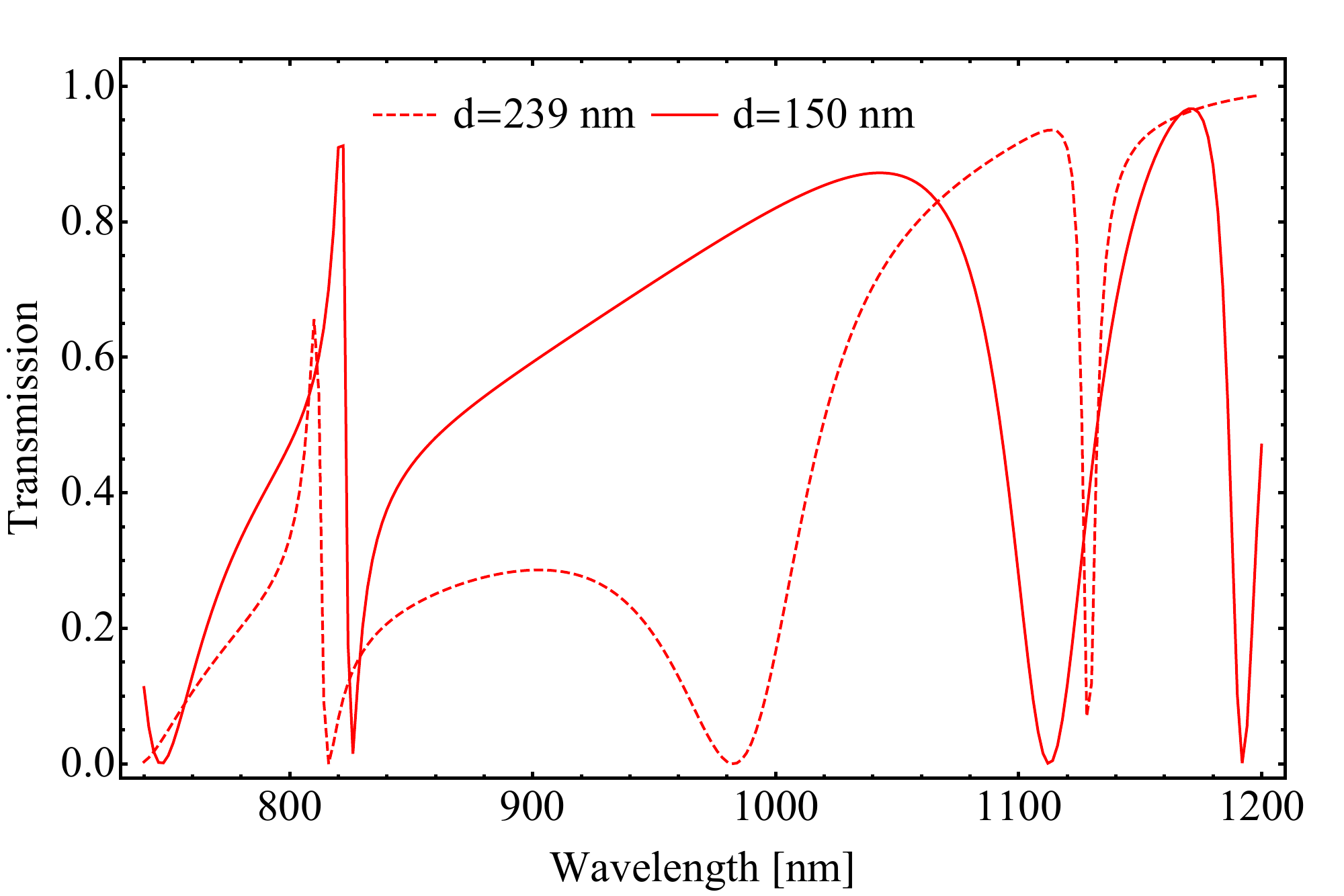}
\includegraphics[width=0.49\textwidth]{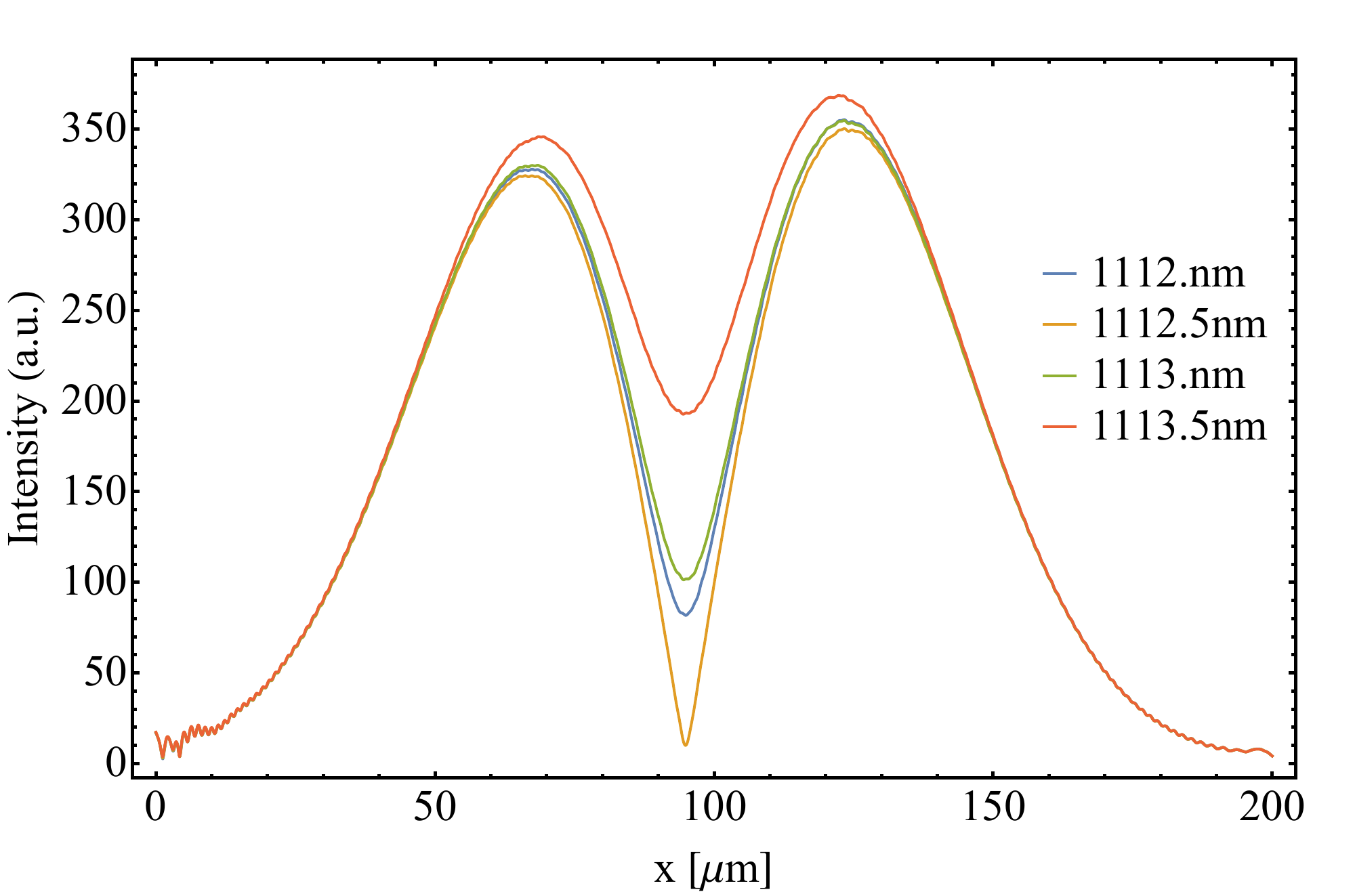}
\caption{Left: RCWA-simulated spectrum of $150$ nm-deep grating illuminated by TE-polarized light impinging at a 3.6$^\circ$ oblique incidence (plain). The dashed line shows as a reference the corresponding spectrum for the $239$ nm-deep grating of the main text. Right: FEM-simulated transmitted beam intensity profiles for wavelengths close to the TE$_0$ even-superposition resonance.}
\label{fig:d150}
\end{figure}

To assess the effect of a nearby higher-order resonance on the first-order spatial differentiation at oblique incidence we simulated in Fig.~\ref{fig:d150} the response of a grating with the same parameters as those used in Sec.~II of the main, but with a finger depth of $d=150$~nm and illuminated by TE-polarized light impinging at a 3.6$^\circ$ oblique incidence. As expected from such a shallower grating, the TE$_0$ even- and odd-superposition resonances occur at 1112 and 1192~nm, respectively. These are more spectrally separated from the TE$_1$ resonances at 747 and 826~nm than for the $239$~nm-deep grating of the main text (reproduced in Fig.~\ref{fig:d150} by the dashed curve). As a consequence, the transmitted beam intensity profiles around the 1112~nm resonance show a higher degree of symmetry than those of Fig.~4 of the main text.

\end{document}